\documentclass[aps,prc,showpacs]{revtex4}
\usepackage{graphicx}
\begin{document}

\title{Quark recombination and elliptic flow}
\author{Scott Pratt}
\email{pratts@pa.msu.edu}
\affiliation{Department of Physics and Astronomy, Michigan State
University, East Lansing, Michigan 48824}
\author{Subrata Pal}
\email{pal@nscl.msu.edu}
\affiliation{National Superconducting Cyclotron Laboratory
and Department of Physics and Astronomy,
Michigan State University, East Lansing, Michigan 48824}

\begin{abstract}
Elliptic flow systematics for different hadron species have been
explained by quark coalescence models. It has been argued that
the elliptic asymmetry $v_2$ should scale with the number of quarks that
comprise the hadron. We show how these arguments are sensitive to the relative
role of asymmetries in the phase space density vs. asymmetries in the effective
volume of emission. We also discuss the degree to which coalescence arguments
differ from thermal models.  Illustrative calculations based on solving the
Boltzmann equation are presented along with the results of blast-wave
models. Although the issue is complicated, ambiguities might be
clarified by measurements of source-size parameters for nucleons at higher
transverse momenta.
\bigskip
\end{abstract}

\maketitle

\section{Introduction}
\label{sec:intro}

One of the surprising early measurements from RHIC shows that the proton/pion
ratio reaches or even exceeds unity for transverse momenta $p_t$ above 2 GeV/c
\cite{poverpidata}. One explanation for this phenomena is that quarks
originating from different nucleon-nucleon collisions recombine via coalescence
mechanisms \cite{voloshin,muellergroup,kogroup}. If this is indeed the
explanation, it will make a case for having created a new state of matter where
memory of the original color singlet hadrons is lost as the constituent partons
become deconfined before going through a ``mix-and-match'' process, referred to
as quark recombination.

Coalescence arguments have also been invoked to explain systematics of elliptic
flow asymmetries \cite{voloshin,muellergroup,friesasymmetry,kogroup,molnar}.
These analyses are based on the parameter \cite{ollit,volozh},
\begin{equation}
\label{eq:v2def}
v_2(p_t)\equiv \frac{\int  d\phi \left(dN/d\phi dp_t\right) \cos 2\phi}
{\int d\phi \left(dN/d\phi dp_t\right) },
\end{equation}
where the azimuthal angle $\phi$ is measured relative to the reaction plane. 
It is argued that if the underlying quarks have an asymmetry $v_2^{(q)}$, 
hadrons will have an asymmetry \cite{voloshin,molnar,friesasymmetry},
\begin{equation}
\label{eq:v2scaling}
v_2^{(n_q)}(p_t)=n_q v_2^{(q)}(p_t/n_q),
\end{equation}
where $n_q$ is the number of quarks comprising the hadron. This might
explain the fact that baryons seem to have $\sim 50\%$ higher asymmetries than
mesons \cite{expv2scaling}.

The main purpose of this paper is to investigate the validity of the scaling
relation for $v_2$ of Eq. (\ref{eq:v2scaling}). Aside from the usual caveats
concerning coalescence, Eq. (\ref{eq:v2scaling}) requires an assumption that
the average phase space density has the same asymmetry as the invariant
spectra. After demonstrating how this assumption plays a role  we
present sample calculations to illustrate the circumstances under which
Eq. (\ref{eq:v2scaling}) is valid. For blast-wave models of the emitting state
\cite{schn,huovinen,csorgo,lisaretiere}, we will show that the validity is
sensitive to seemingly arcane choices of how to parameterize the
model. Boltzmann equations are also solved for the partonic evolution where
validity of Eq. (\ref{eq:v2scaling}) is found to be sensitive to the initial
conditions.

The next section presents a discussion of quark recombination through
coalescence with an emphasis on explaining the subtle differences between
coalescence and thermal models. The subsequent sections present results from
the blast-wave and Boltzmann calculations respectively. Results are summarized
in Sec. \ref{sec:summary}.

\section{Theory: Coalescence and Elliptic Flow}
\label{sec:theory}

Coalescence arguments have been applied in numerous instances
\cite{schwarz,gyulassy,dani,nagle,kahana,mattiello,schaffner} to problems of
fragment production in nuclear physics. It has been especially successful in
describing the systematics of light nuclei production in high energy heavy-ion
collisions where a high degree of thermalization is attained and temperatures
far exceed typical nuclear binding energies. Coalescence shares many features
of thermal models, and in fact provides the same results in the limit of high
temperature relative to the binding energy \cite{mekjian}. Although the focus
of this paper is quark coalescence, the theory discussed in this section can be
equally well applied to the coalescence of nucleons to form deuterons or light
clusters. The main theoretical difference with quark coalescence is that one
must be more cognizant of the effects of the non-zero binding energy and of
relativistic motion. Whereas the deuteron binding energy is only 2.2 MeV and
the momentum components of the nucleon-nucleon relative wave function has most
of its strength in the range of tens of MeV/c, constituent quarks may form
hadrons whose mass differs from the sum of the constituent quark masses by many
hundreds of MeV. Momentum components of the quark's relative motion also tend
to be in the range of hundreds of MeV/c. Furthermore, one must add the
qualifier that hadrons cannot necessarily be expressed in terms of constituent
quarks, which ignores the gluons, sea quarks or vacuum fluctuations of meson
fields.

In this section we further outline some of the subtle issues mentioned above,
but rather than focusing on these issues, we will mainly consider a simplified
case where $n_q$ constituent quarks each of mass $m_q$ combine to form a hadron
of mass $n_qm_q$. This allows us to sidestep some of the more delicate points
about coalescence and derive some simple relations for the behavior of elliptic
flow as a function of the number of quarks.

\subsection{Review of Coalescence Theory}

The meaning of the term coalescence varies somewhat throughout the literature
\cite{schwarz,gyulassy,dani,nagle,kahana,mattiello,schaffner}. For the purposes
of our discussion, we assume that it refers to the formation of a bound state
of two pre-existing constituents via the sudden approximation, i.e., at the
moment of coalescence the binding interaction is turned on suddenly and the
probability that constituents $a$ and $b$ will form a composite object $C$ is
\begin{equation}
\label{eq:coalescencemaster}
f_C({\bf P}_C,{\bf R},t_c)=\int d^3r d^3q \: W_{ab}^{(C)}({\bf q},{\bf r}) \:
f_a(m_a{\bf P}_C/(m_a+m_b)+{\bf q},{\bf R}+{\bf r}/2,t_c)
f_b(m_b{\bf P}_C/(m_a+m_b)-{\bf q},{\bf R}-{\bf r}/2,t_c) .
\end{equation}
Here, the phase space densities are denoted by $f_i$ and $W$ refers to the
Wigner decomposition of the relative squared wave function. ${\bf P}_C$ is the
momentum of the composite object, $t_c$ the coalescence time, and $\bf R$ and
$\bf r$ are the center-of-mass and relative coordinates.  Although this
expression is non-relativistic, the problem can be considered in a frame where
the velocity of the constituent particle $V$ is zero or small.

If the characteristic spread of the relative momenta and relative coordinates
in $W({\bf q},{\bf r})$ are small compared to the scale of the initial momenta 
and spatial sizes, $W$ may be replaced with a delta function,
\begin{equation}
\label{eq:coalapprox}
W_{ab}^{(C)}({\bf q},{\bf r}) \approx\delta({\bf r})\delta({\bf q}).
\end{equation}
The simplified coalescence formula is then,
\begin{equation}
\label{eq:coalescence}
f_C({\bf P}_C,{\bf R},t_c)\approx f_a(m_a{\bf P}_C/(m_a+m_b),{\bf R},t_c) \:
f_b(m_b{\bf P}_C/(m_a+m_b),{\bf R},t_c).
\end{equation}

For a thermal source, we consider a Boltzmann form for the phase space
density,
\begin{equation}
\label{eq:thermal}
f_i(m_i{\bf V},{\bf R},t_c) = \exp
\left[-(E^\prime_i-\mu_i({\bf R},t_c))/T({\bf R},t_c) \right],
\end{equation}
where $E'$ refers to the energy in the frame of the thermalizing
medium. Inserting the thermal forms for $f_a$ and $f_b$ in
Eq. (\ref{eq:coalescence}) nearly reproduces the thermal form for $f_C$ with
$\mu_C=\mu_a+\mu_b$. It differs by the factor $\exp(-B'/T)$ where $B'$ is the
binding energy measured in the frame of the thermalized medium,
\begin{equation}
B'=E'_a+E'_b-E'_C .
\end{equation}
For the coalescence of nucleons into deuterons, the binding energy of 2.2 MeV
is much smaller than the temperature, making the coalescence and thermal forms
nearly indistinguishable.

The source-function formalism, which is commonly used in two-particle
correlations \cite{llope}, can also be applied to this problem. In this
formalism the last scatterings with third bodies are treated as randomizing
interactions which can place the particles into bound states. The source
functions are defined in terms of the quantum $T$ matrices which describe the
scattering from third bodies into the final two-body state. The momentum
distribution of a composite object created from two particles,
\begin{equation}
\label{eq:tmatrix}
\frac{dN^{(C)}}{d^3p_C}=\sum_{f'}
\left|\int d^4x_a d^4x_b \: T_{f'}(x_a,x_b) U(x_a,x_b;p_C)\right|^2,
\end{equation}
where $U(x_a,x_b;p_C)$ is the evolution matrix for particles evolving from
space-time points $x_a$ and $x_b$ into the asymptotic state of a bound
particle, $C$, with momentum $p_C$. The remainder of the system will evolve
into the state $f'$. By imposing the condition that the sources are
independent, (apart from the final-state interaction), the $T$ matrices can be
expressed as a product,
\begin{equation}
T_{f'}(x_a,x_b)\rightarrow
T_{f'_a}(x_a)T^*_{f'_b}(x_b),~~~\sum_{f'}\rightarrow \sum_{f'_a,f'_b},
\end{equation}
where $f'_a$ and $f'_b$ refer to the two independent sources of the particles
$a$ and $b$. Inserting the factorized form into Eq. (\ref{eq:tmatrix}),
\begin{eqnarray}
\frac{dN^{(C)}}{d^3p_C}&=&\sum_{f_a,f_b}
\left|\int d^4x_a d^4x_b \: T_{f_a}(x_a)T_{f_b}(x_b) U(x_a,x_b;p_C)\right|^2 
\nonumber\\
&=&\sum_{f_a,f_b}\left|\int d^4x_a d^4x_b T_{f_a}(x_a)T_{f_b}(x_b)
\exp\left[i(p_a+p_b)\cdot(m_ax_a+m_bx_b)/(m_a+m_b)\right]
u_C(x'_a-x'_b)\right|^2 ,
\end{eqnarray}
where the two scattering centers evolve into states $f'_a$ and $f'_b$.  The
center-of-mass behavior has been factored out of the evolution matrix, and $u$
is the evolution matrix in the frame where ${\bf P}_C=0$, with $X'_a-X'_b$
referring to the coordinates measured in that frame.  (If the sources are not
independent, there would have been correlations in the emission of $a$ and $b$
in the absence of final-state interactions.)

Invoking properties of Fourier transform,
\begin{eqnarray}
\label{eq:sourcemess}
\frac{dN^{(C)}}{d^3P_C}&=&\frac{1}{(2\pi)^4}\int d^4X_a d^4X_b d^4r d^4q \:
S_a(m_aP_C/(m_a+m_b)+q,X_a) S_b(m_bP_C/(m_a+m_b)-q,X_b)
e^{-iq\cdot r} \nonumber\\
&&~~~~~~~~ \times u^*(X'_a-X'_b+r'/2) u(X'_a-X'_b-r'/2) ; \nonumber\\
S_i(p_i,X)&\equiv& \sum_{f_a}\int d^4x T_{f_i}^*(X+x/2)T_{f_i}(X-x/2)
e^{-ip_i\cdot X}.
\end{eqnarray}
When the source function $S_i(p,x)$ is evaluated on-shell, $p_0=E_p$, it can be
identified with the probability of emitting a particle of type $i$ with
momentum ${\bf p}$ from space-time point $x$.

A more tractable expression can be obtained by applying the smoothness
approximation \cite{prattsmoothness,wiedemannsmoothness},
\begin{equation}
S_a(m_aP_C/(m_a+m_b)+q,X_a) S_b(m_bP_C/(m_a+m_b)-q,X_b)\approx
S_a(m_aP_C/(m_a+m_b),X_a) S_b(m_bP_C/(m_a+m_b),X_b).
\end{equation}
This permits changing the integral over $q$ into a delta function which sets
$r'$ to zero. Equation (\ref{eq:sourcemess}) can then be written as
\begin{equation}
\label{eq:sourcemess2}
\frac{dN^{(C)}}{d^3P_C}=\int d^4X_a d^4X_b \:
S_a(m_aP_C/(m_a+m_b),X_a) S_b(m_bP_C/(m_a+m_b),X_b) |u_C(X'_a-X_b')|^2 .
\end{equation}
This approximation is exact for thermal sources, since the Boltzmann factor is
only a function of $P_a+P_b$, and is independent of the relative momentum.

Unfortunately, the evolution matrix $u_C(X'_a-X'_b)$ involves a time
difference.  If the two times $t'_a$ and $t'_b$ are equal, $u$ can be
replaced with the relative wave function which is often a well-understood
object. A third approximation can then be made by introducing time independent 
evolution matrix $\phi$, i.e.,
\begin{equation}
\label{eq:equaltime}
|u_C(X'_a-X'_b)|^2\approx (2\pi)^3|\phi_C({\bf X}'_a-{\bf X}'_b)|^2.
\end{equation}
Without this last approximation it would have been essential to understand the
details of the evolution of the particles for times between the emissions of
the first and second particle. Since the particles are, by definition, not
moving rapidly in this frame, the wave function at this time should not change
substantially and this should be a reasonable approximation. Putting the three
approximations together, i.e., source factorization, smoothness, and ignoring
of the offsets of emission times, we obtain an expression similar to that used
in correlation studies
\begin{equation}
\label{eq:sourcemaster}
\frac{dN^{(C)}}{d^3P_C}=(2\pi)^3 \int d^4X_a d^4X_b \:
S_a(m_aP_C/(m_a+m_b),X_a) S_b(m_bP_C/(m_a+m_b),X_b)
|\phi_C({\bf X}'_a-{\bf X}'_b)|^2 .
\end{equation}

After approximating the squared relative wave function as a $\delta$ function,
and noting that the phase space density can be related to the source function
by the relation,
\begin{equation}
\label{eq:coalvssource}
f_a(m_a{\bf P}_C/(m_a+m_b),{\bf x},t_c) =
(2\pi)^3 \int d^4X_a \: S_a(E_a,m_a {\bf P}_C/(m_a+m_b),X_a)
\delta({\bf x}-{\bf X}-{\bf v}(t_c-X_0)),
\end{equation}
the coalescence formula of Eq. (\ref{eq:coalescence}) would be reproduced,
except that the source function in Eq. (\ref{eq:sourcemaster}) is evaluated
off-shell.
\begin{equation}
m_aE_C/(m_a+m_b)\ne E_a(m_a{\bf P}_C/(m_a+m_b)).
\end{equation}
Within the smoothness approximation, there is no difference in replacing the
energy arguments of the source functions in Eq. (\ref{eq:sourcemaster}) with
on-shell values.

For a thermal source, the product of the source functions becomes a Boltzmann
factor, $\exp(-E_C/T)$, and hence Eq. (\ref{eq:sourcemaster}) is equivalent to
a thermal expression. Thus, the coalescence expression,
Eq. (\ref{eq:coalescencemaster}), the thermal expression,
Eq. (\ref{eq:thermal}), and the source function expression,
Eq. (\ref{eq:sourcemaster}), become equivalent for thermal distributions when
the binding energy is zero. The lack of dependence on the binding energy in
coalescence formulas is due to the sudden approximation, which presumes that
the binding potential is turned on suddenly, i.e., energy is not
conserved. From a quantum perspective, the source function in
Eq. (\ref{eq:coalvssource}) is found by factoring quantum matrix elements with
the density of states. If the matrix elements are assumed to be featureless,
and if the density of states of the sources behaves as $\exp(E/T)$, a thermal
form is then obtained. When forming hadrons, it is hard to argue that the
interactions with third bodies do not provide some additional leverage for
creating final states with larger binding energies, i.e., the phase space
density of protons should be greater than that of deltas. As can be garnered
from the list of approximations above, neither the coalescence expression nor
the thermal expression can be justified to better than the 10\%
level. Moreover, once a decision has been made regarding the choice of one of
the formalisms, additional choices must be made for parameters such as the
temperature and rest frame of the thermal source. For the coalescence formula
it is necessary to make a choice of a rest frame for the coalescence. In
this Lorentz frame the interaction between $a$ and $b$ turns on
instantaneously. Momentum is conserved, but energy is not conserved. In all
other frames, neither energy nor momentum are conserved. Again, it should be
stressed that these subtle sensitivities to parameters and choice of reference
frames disappear when the binding energy is small.

To illustrate the sensitivity of coalescence pictures to binding energy, let us
consider the extreme case of two constituent quarks of mass 350 MeV
coalescing to form a pion. If the quarks are to coalesce in the two-particle
rest frame, the energy will change by 60\% from the coalescence in that
frame. If the coalescence is observed in the lab frame, the pair will 
loose 60\% of its momentum in the coalescence process. Whereas,
if the coalescence frame is the laboratory frame, the laboratory momentum of
the two quarks will be conserved in this process. Although, pions are an
extreme example, most coalescence pictures involve mass changes at the
20\% level, which provides a feel for the confidence with which one can apply
coalescence prescriptions.

Whether recombination is thermal or sudden by nature, the two expressions share
an important property. Both formulas presume that the $a$ and $b$ components
exist independently of one another. This will not be true if the partons, for
example, originate from the same jet. It is the validity of the factorization
that is related to whether or not hadronization can be described as the
recombination of partons without regard to whether the two owe their existence
to the same initial nucleon-nucleon collision. For many of the calculations
presented in the following sections, the issue of thermal vs. sudden will be
side-stepped by focusing on the coalescence of two quarks with constituent
masses $m_q$ which combine to make a meson of mass $2m_q$.

\subsection{Coalescence and Elliptic Flow}

In this subsection, and in the examples presented in the next section, we
consider the coalescence of $n_q$ quarks into a hadron of mass $M_C=n_qm_q$,
where $m_q$ is the constituent quark mass. Thus, we ignore many of the subtle
issues concerning the binding energy, and the relative merits of one formalism
with another. The expression analogous to Eq. (\ref{eq:coalescence}) for the
coalescence of $n_q$ quarks is
\begin{equation}
f_C({\bf P},{\bf R},t) \approx
\left[f_q(m_q{\bf P}/M_C,{\bf R},t)\right]^{n_q} .
\end{equation}
The hadron spectra is found by integrating the phase space density over
coordinate space,
\begin{eqnarray}
\label{eq:spectra}
\frac{dN_C}{d^3P}&=& \frac{(2S_C+1)}{(2\pi)^3}
\int d^3r f_C({\bf P},{\bf r},t) \nonumber\\
&=& \frac{(2S_C+1)}{(2\pi)^3} \int d^3r
\left[f_q({\bf P}/n_q,{\bf R},t)\right]^{n_q},
\end{eqnarray}
where $S_C$ is the spin of the composite particle $C$.

The principal goal of this paper is to investigate the behavior of the angular
asymmetry which is quantified by the parameter $v_2$ defined in
Eq. (\ref{eq:v2def}). If the composite particle is made up of $n_q$ quarks,
each with the same phase space density $f_q$, we have from Eqs.
(\ref{eq:v2def}) and (\ref{eq:spectra})
\begin{equation}
v_2^{(n_q)}(p_t)=\frac{\int d\phi d^3r
\left[f_q(p_t/n_q,\phi,{\bf r},t_c)\right]^{n_q}\cos 2\phi}
{\int d\phi d^3r \left[f_q(p_t/n_q,\phi,{\bf r},t_c)\right]^{n_q}},
\end{equation}
where $\phi$ refers to the azimuthal angle of the momentum.

The phase space density can be written as a product of the spectra and an
effective density, $\rho_q({\bf r},y_t,\phi)$, defined by the relation,
\begin{equation}
\rho_q({\bf r},y_t,\phi)\equiv \frac{f_q(y_t,\phi,{\bf r})}
{(2\pi)^3(dN^{(q)}/d^3p)}.
\end{equation}
Here, $\rho_q$ is non-zero over the region where particles are emitted. If the
phase space density is constant within a fixed volume, $\Omega$, then 
$\rho_q = 1/\Omega$. If $\rho_q$ is independent of $\phi$, $v_2^{(C)}$ can be
written in terms of the quark spectra,
\begin{equation}
\label{eq:friesapprox}
v_2^{(n_q)}(y_t)=\frac{\int d\phi \left[dN^{(q)}/d^3p\right]^{n_q}\cos 2\phi}
{\int d\phi \left[dN^{(q)}/d^3p\right]^{n_q}},
\end{equation}
where $dN^{(q)}/d^3p$ is evaluated for quarks at a fixed transverse rapidity,
$y_t$. If the quark spectra has an elliptic asymmetry as
$dN^{(q)}/d^3p \propto (1+\epsilon\cos 2\phi)$, then $v_2^{(q)}=\epsilon/2$
and $v_2^{(n_q)}$ can be expressed in terms of $v_2^{(q)}$,
\begin{eqnarray}
\label{eq:scaling_master}
v_2^{(n_q)} &=& \frac{\int d\phi \:
\left[1+2v_2^{(q)}\cos 2\phi\right]^{n_q} \cos 2\phi}
{\int d\phi \left[1+2v_2^{(q)}\cos 2\phi\right]^{n_q}} \nonumber\\
&=&\frac{1}{2}\frac{\sum_{n=1,3,\cdots}^{n_q} (n+1)
\left(v_2^{(q)}\right)^n
\frac{n_q!}{(n_q-n)!\left([(n+1)/2]!\right)^2}}
{\sum_{n=0,2,\cdots}^{n_q} \left(v_2^{(q)}\right)^n
\frac{n_q!}{(n_q-n)!\left[(n/2)!\right]^2}}.
\end{eqnarray}
For $n_q=2,3$, the expression becomes \cite{molnar}
\begin{eqnarray}
\label{eq:scaling23}
v_2^{(2)}&=&2v_2^{(q)}\frac{1}{1+2(v_2^{(q)})^2} \nonumber\\
v_2^{(3)}&=&3v_2^{(q)}\frac{1+(v_2^{(q)})^2}
{1+6(v_2^{(q)})^2}.
\end{eqnarray}
For small $v_2^{(q)}$ one obtains the simple scaling relation,
Eq. (\ref{eq:v2scaling}).

\begin{figure}
\centerline{\includegraphics[width=0.45\textwidth]{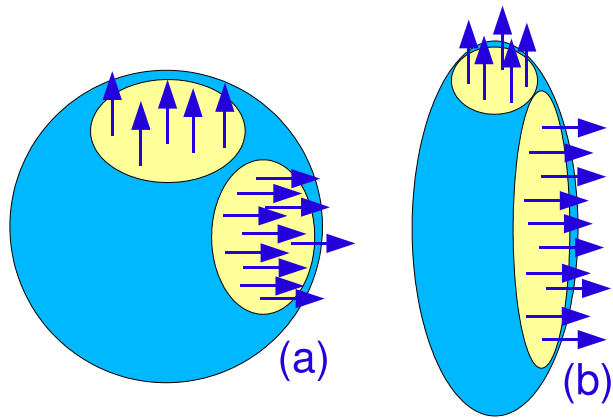}}
\caption{\label{fig:effvolcartoon}
Two sources are illustrated above. The particles of a given velocity are
represented by arrows. In (a) the effective volumes of the right-moving and
upward-moving particles are the same but the right-moving particles have a
higher phase space density. In (b) the phase space densities are the same but
the effective volumes differ. In both cases there are twice as many particles
moving to the right as upward so $v_2$ is identical. But in (a) the $v_2$ for
a bound state of $n_q$ particles will increase by a factor of $n_q$ while 
in (b) the composite particle will have the same $v_2$ as the constituents.}
\end{figure}

It is this simple scaling relation that has been cited as evidence for quark
recombination \cite{voloshin,molnar,friesasymmetry}. However, the validity of
this relation hinges on the assumption that the effective density is
independent of $\phi$. The nature of this assumption can be understood by
considering two examples illustrated in Fig. \ref{fig:effvolcartoon}. In the
left side the effective volumes for emission upwards and to the right are
identical, and the higher number of particles emitted to the right is due to
the higher average phase space density of those particles. Thus, for this
example, the elliptic asymmetry for dimers will be double the elliptic
asymmetry for monomers and will satisfy the scaling relation,
Eq. (\ref{eq:v2scaling}). For the emission illustrated in the right panel of
Fig. \ref{fig:effvolcartoon}, the average phase space density for emission in
the two directions are identical, and the higher number of particles emitted to
the right is due to a larger effective volume. In this case, the asymmetry for
monomers and dimers is the same.

To illustrate the same effect shown in Fig. \ref{fig:effvolcartoon}
algebraically, let us consider a phase space density characterized by a
Gaussian profile,
\begin{equation}
f_q(p_t,\phi,{\bf r})=2^{3/2} \bar{f}(p_t,\phi)
\exp\left[-r^2/(2R^2(p_t,\phi))\right],
\end{equation}
where $\bar{f}$ is the phase space density averaged over the
coordinate space. The probability for emitting a coalesced $n_q$-quark object
is then
\begin{equation}
\frac{dN_C}{d^3p}=\frac{(2S_C+1)2^{3n_q/2}}{(2\pi n_q)^{3/2}}
R^3(p_t,\phi) \left[\bar{f}(p_t,\phi)\right]^{n_q}.
\end{equation}
If $v_2^{(\bar f)}$ and $v_2^{(\Omega)}$ refer to the elliptic asymmetries of
the average phase space density and effective volumes of the single quarks,
\begin{eqnarray}
v_2^{(\bar{f})}(p_t)
&\equiv&\frac{\int d\phi \: \bar{f}(p_t,\phi)\cos 2\phi}
{\int d\phi \: \bar{f}(p_t,\phi)}, \nonumber\\
v_2^{(\Omega)}(p_t)
&\equiv&\frac{\int d\phi \: R^3(p_t,\phi)\cos 2\phi}
{\int d\phi \: R^3(p_t,\phi)} ,
\end{eqnarray}
the asymmetry for the $n_q$-quark object can be written as
\begin{equation}
\label{eq:generalscaling}
v_2^{(n_q)}=v_2^{(\Omega)}+n_qv_2^{(\bar{f})}.
\end{equation}
Equation (\ref{eq:generalscaling}) is again based on the assumption that the
asymmetries are not large, but includes the chance that the asymmetries can
originate from either asymmetries of the effective volume or asymmetries of the
average phase space density.

If the effective volumes are independent of $\phi$, $v_2^{(\Omega)}$ is zero.
The Eqs. (\ref{eq:scaling_master}) and (\ref{eq:generalscaling}) are then identical
and the asymmetry scales linearly with $n_q$. Such a case is illustrated in the
left panel of Fig. \ref{fig:effvolcartoon}. On the other hand, if the average 
phase space density
is independent of $\phi$, $v_2^{(\bar{f})}$ is zero and the asymmetry is then
independent of $n_q$. This will be the case for emissions of the type
illustrated in the right panel of Fig. \ref{fig:effvolcartoon}. Several
examples are explored in the next section, with the purpose of understanding
what drives the asymmetry, the effective volume or the average phase space
density.

The effective volumes, $R^3$ in the Gaussian case, can be determined
experimentally through correlation measurements \cite{starcor,phenixcor,lisa}
or with coalescence measurements \cite{llope,stardbar,palpratt}. This volume
can be defined in terms of the phase space density,
\begin{equation}
\label{eq:omegadef}
\Omega(y_t,\phi)\equiv\frac{\left[\int d^3r f_q(y_t,{\bf r})\right]^2}
{\int d^3r [f_q(y_t,{\bf r})]^2}.
\end{equation}
For a Gaussian source of dimensions $R_x$, $R_y$ and $R_z$,
$\Omega=2^{3/2}R_xR_yR_z$. In considering the coalescence of nucleons,
rather than that of quarks, the volume may be determined with
two-nucleon correlation measurements. The effective volume may be also
estimated from coalescence ratios. The coalescence volume can be defined by the
ratio of spectra of two particles $a$ and $b$ that combine to form a coalesced
object $C$ as can be seen in Eq. (\ref{eq:omegadef}),
\begin{equation}
\Omega=\frac{(2S_C+1)}{(2S_a+1)(2S_b+1)}
\frac{(dN_a/d^3p)(dN_b/d^3p)}{dN_C/d^3P}.
\end{equation}
For small asymmetries, this yields $v_2^{(\Omega)}=v_2^{(a)}+v_2^{(b)}
-v_2^{(C)}$, which is basically a tautological restatement of
Eq. (\ref{eq:generalscaling}). Other methods for measuring the source size as a
function of the azimuthal angle, will provide independent means for
determining whether the $v_2^{(\Omega)}$ is independent of $\phi$. Examples of
such methods are identical pion correlations or proton-lambda correlations.

\section{Blast Wave Models}

Blast wave descriptions \cite{schn,huovinen,csorgo,lisaretiere}, provide a
simple means by which one can test thermal concepts along with those for
collective flow.  Typically, a blast-wave model requires four parameters: a
temperature $T$, a breakup time $\tau$, a transverse radius $R$ and a maximum
transverse collective rapidity, $\rho_0$. When considering elliptic
asymmetries, one may also use two parameters for the transverse radii, $R_x$
and $R_y$, and two parameters to describe the transverse collective motion.
There are many variants of blast-wave models that differ from the form by which
the space-time distributions are parameterized. Of the three parameterizations
considered here, each assumes that the distribution of thermal sources is
invariant to boosts along the beam directions. Although the choice of the form
seems somewhat arbitrary when analyzing spectra, we find that the different
forms result in rather different behavior in the elliptic flow.

\subsection{Shell parameterization for the blast-wave}

Asymmetries can be added to a thermal blast-wave in four independent ways via
the temperature, the collective velocity, the chemical potential, or the
volume.  To compare the effects of these four types of asymmetries, we consider
a simple azimuthally symmetric shell
\begin{equation}
\label{eq:shell0}
\frac{dN}{d\phi_p dp_t} \sim \int d\phi_u d\eta \: \cosh\eta \: \rho_s(\phi_u)
\exp\!\left[ \frac{-\cosh\eta \: m_t\sqrt{1+u_\perp^2(\phi(s))}
+ u_\perp(\phi_u) p_t\cos(\phi_p-\phi_u)} {T(\phi_u)+n\mu(\phi_u)/T(\phi_u)}
\right].
\end{equation}
Here $u_\perp$ is the transverse collective velocity, $\rho(\phi_u)$ is the
angular density for sources having a collective flow in the radial direction
$\phi_u$ and $\eta$ is the longitudinal rapidity of the source. The four
asymmetries can be parameterized in the following way,
\begin{eqnarray}
\rho(\phi_u)&=&(1+\epsilon_V\cos2\phi_u)\rho, \nonumber\\
\exp\!\left[\mu(\phi_u)/T(\phi_u)\right] &=& 
(1+\epsilon_\mu\cos 2\phi_u)\exp(n_q\mu/T), \nonumber\\
T(\phi_u)&=&(1+\epsilon_T\cos 2\phi_u)T, \nonumber\\
u_\perp(\phi_u)&=&(1+\epsilon_u\cos 2\phi_u)u_\perp.
\end{eqnarray}
Inserting these equations into (\ref{eq:shell0}), the expressions are then
expanded to lowest power in $\epsilon$ which allows the integrals to be
performed analytically. The resulting expressions for each of the four
expansion are:
\begin{eqnarray}
v_{2,V}&=&\epsilon_V\frac{J_V}{J_0}, \nonumber\\
v_{2,\mu}&=&n_q\epsilon_\mu \frac{J_V}{J_0}, \nonumber\\
v_{2,T}&=&\epsilon_T\left( \frac{m_t\sqrt{1+u_\perp^2}}{T}J_E
-\frac{u_\perp p_t}{T}J_p\right)\frac{1}{J_0}, \nonumber\\
v_{2,u}&=&\epsilon_u\left(-\frac{u_\perp^2m_t}{T\sqrt{1+u_\perp^2}}J_E
+\frac{u_\perp p_t}{T}J_p\right)\frac{1}{J_0},
\end{eqnarray}
where $J_0$, $J_E$ and $J_p$ are simple combinations of Bessel functions,
\begin{eqnarray}
J_0&=&2K_1(z_E)I_0(z_p), \nonumber\\
J_V&=&K_1(z_E)I_2(z_p), \nonumber\\
J_p&=&\frac{1}{2}K_1(z_E)\left[I_1(z_p)+I_3(z_p)\right], \nonumber\\
J_E&=&\left[K_0(z_E)+K_1(z_E)/z_E\right]I_2(z_p),
\end{eqnarray}
with $z_E=(m_t/T)\sqrt{1+u_\perp^2}$ and $z_p=u_\perp p_t/T$.

\begin{figure}
\centerline{\includegraphics[width=0.4\textwidth]{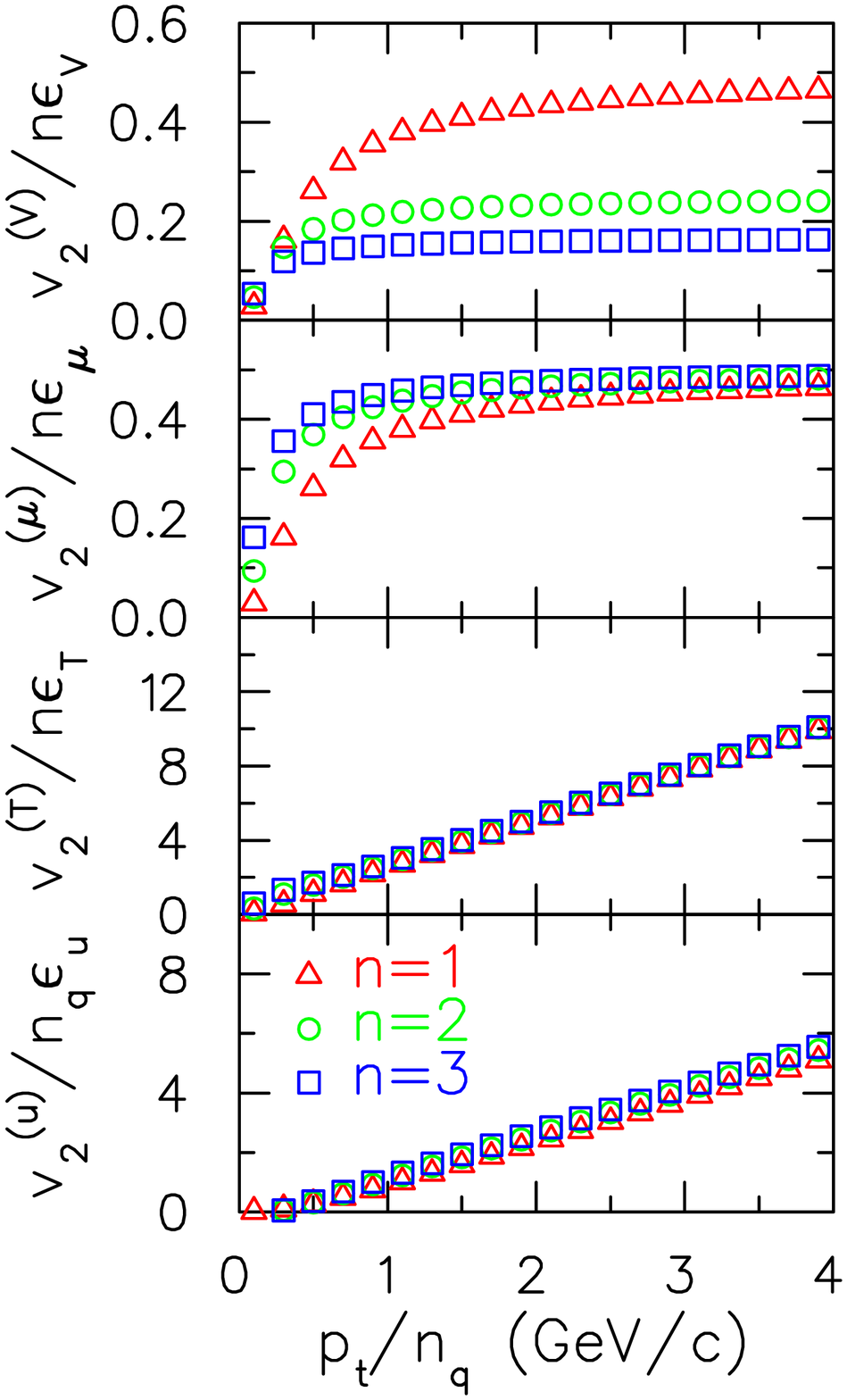}}
\caption{\label{fig:shell} Scaled values for elliptic flow as a function of
scaled transverse momentum for constituent quarks each of mass $m_q=350$ MeV
coalescing to form composite objects of $n_q=1,2,3$ quarks with mass
$M_C=n_qm_q$.  Asymmetries are introduced into the cylindrical-shell
parameterization of the blast-wave by adding elliptic distortions to the
thickness of the envelope (upper panel), the density as driven by a chemical
potential (second panel), the temperature (third panel) and the collective
velocity (lower panel). Scaling is significantly violated for the upper two
panels.}
\end{figure}

Figure \ref{fig:shell} displays the resulting values of $v_2$ for each of the
four asymmetries as a function of the scaled transverse momentum. For these
calculations, the masses of the coalesced hadrons are assumed to be $n_q\times$
350 MeV. By plotting $v_2/n_q$ vs. the $p_t/n_q$, satisfaction of quark-number
scaling of Eq. (\ref{eq:v2scaling}) will provide indistinguishable lines for
different values of $n_q$. The upper panel shows asymmetries from adding an
elliptic distortion to $\rho(\phi_u)$. This type of asymmetry will result from
distorting the shape of the shell while keeping the magnitude of the collective
velocity fixed, or by altering the thickness of the shell. Effectively, these
distortions represent an elliptic asymmetry to the effective volumes of the
source as described in Fig. \ref{fig:effvolcartoon}. Scaling is then strongly
violated since $v_2$ is approximately independent of $n_q$. In the second
panel, results are shown for adding an elliptic asymmetry to the chemical
potential. This is same as in the upper panel, except that it scales with an
extra power of $n_q$ since the density enters as $\exp(\mu/T)$.  Quark-number
scaling then appears to be well satisfied except at low $p_t$.

The lower two panels of Fig. \ref{fig:shell} show the result of elliptic
distortions to the temperature $T$ and to the magnitude of the collective
velocity $\bar u$. In both cases, quark-number scaling appears to be very well
satisfied. But, in these cases, satisfaction of quark-number scaling is driven
by the fact that $v_2$ rises linearly with $p_t$. If mass effects are
neglected, $v_2$ can only depend on $p_t$ unless the chemical potential is
varied, which will introduce an $n_q$ dependence. If $v_2$ rises linearly,
scaling both the $x$ and $y$ axes by $1/n_q$ will yield a line with the same
slope, and quark-number scaling should automatically be satisfied. In fact,
universal curves will result if scaled by any number, but not just the quark
number. It should be pointed out that the linear behavior with $p_t$ does not
hold at low $p_t$, where $v_2$ must be quadratic. The region over which it
appears quadratic depends on the collective velocity $\bar u $. For smaller
$\bar u $ the quadratic region becomes noticeably larger.

Even though the asymmetries in temperature and collective velocity produce
excellent linear results for $v_2$ vs. $p_t$, the superposition of several
shells with different values of $\bar u $ and $T$ can lead to non-linear
behavior for $v_2$ vs. $p_t$. This is because the slope can depend on the
parameters, and by changing $p_t$, the relative contribution of one shell may
increase with respect to another. When $v_2$ vs. $p_t$ is not linear,
quark-number scaling is more significant.

\subsection{Blast-Wave Parameterization of Retiere and Lisa}

The next parameterization we present was used to analyze elliptic flow and
correlations in \cite{lisaretiere}. This model describes a uniform transverse
density profile that is cut off by an elliptic surface. 
The normalized elliptical radius $\tilde{r}$ corresponding to a surface of 
constant $\tilde{r}$ is defined as
\begin{equation}
\tilde{r}=\sqrt{\frac{x^2}{R_x^2}+\frac{y^2}{R_y^2}}.
\end{equation}
Emission is confined to the region where $\tilde{r}<1$. The transverse rapidity
is defined to rise linearly from the origin,
\begin{equation}
y_{t}=\tilde{r}(\rho_0+\rho_a\cos2\phi),
\end{equation}
where the two parameters, $\rho_0$ and $\rho_a$ parameterize the collective
flow with $\rho_a$ driving the asymmetry. The transverse velocity is then
$u_\perp=\sinh(y_t)$. The direction of the collective flow is chosen to be {\it
perpendicular} to the surfaces of constant $\tilde{r}$.

\begin{figure}
\centerline{\includegraphics[width=0.45\textwidth]{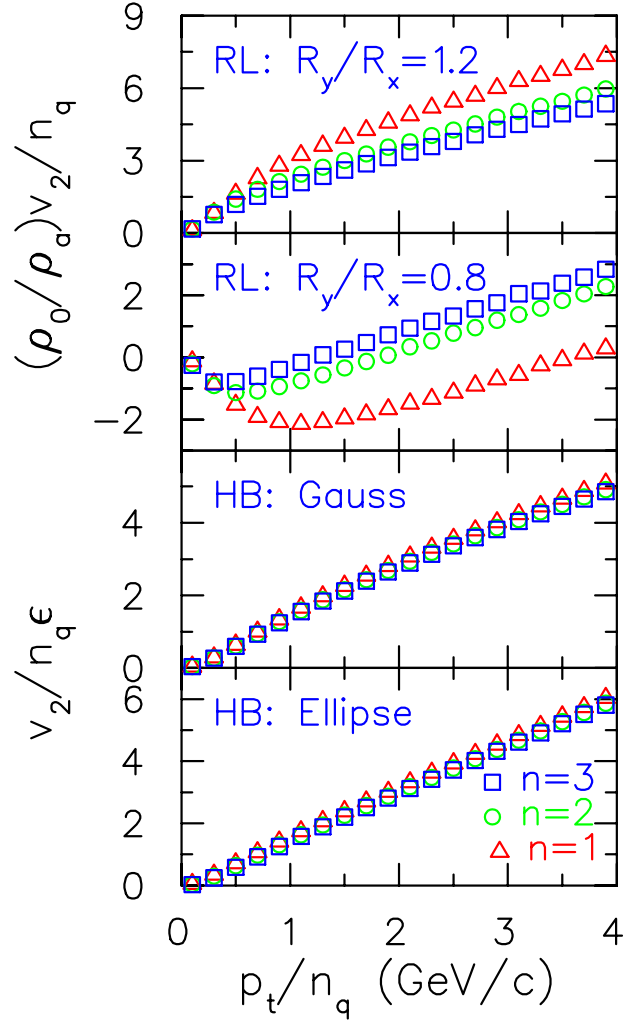}}
\caption{\label{fig:blast} 
Scaled values for elliptic flow as a function of scaled transverse 
momentum for constituent quarks each of mass $m_q=350$ MeV coalescing to form 
composite objects of $n_q=1,2,3$ quarks with mass $M_C=n_qm_q$. A constant 
temperature of $T=100$ MeV is considered for all the calculations. 
The results are in the blast-wave model of Retiere and Lisa \cite{lisaretiere} 
(upper two panels), and to a more hydrodynamically consistent parameterization 
(lower two panels). Quark-number scaling is better realized
by the hydrodynamic parameterization. For the Retiere-Lisa parameterization,
scaling is strongly violated when the ellipse becomes elongated in the in-plane
direction. }
\end{figure}

The upper two panels of Fig. \ref{fig:blast} show $v_2$ from this model with
the parameters, $T=100$ MeV, $\rho_0=0.9$ and $\rho_a=0.1$. The calculations
are for constituent quarks of mass $m_q=350$ MeV and with hadrons of mass
$n_qm_q$. The results are independent of the proper time $\tau$ and are not
independently sensitive to $R_x$ and $R_y$, but instead depend only on the
ratio, $R_y/R_x$. If the matter falls apart quickly after the initial overlap
of the colliding nuclei, then $R_y/R_x>1$.  The upper panel of
Fig. \ref{fig:blast} with $R_y/R_x=1.2$ displays results for such a
scenario. Here $v_2$ of the composite hadrons is calculated using the phase
space density as described in Eq. (\ref{eq:spectra}), and then $v_2$ is divided
by the number of constituent quarks. The scaling criteria of
Eq. (\ref{eq:v2scaling}) is seen to be satisfied rather well in this case as
all the curves lie close to one another.

At later times, the unequal transverse expansion will overcome the initial
out-of-plane extended profile and reach an in-plane-extended geometry.  The
second panel of Fig. \ref{fig:blast} displays the results for
$R_y/R_x=0.8$. In this case, quark-number scaling is strongly
violated. Moreover, $v_2$ becomes negative in some instances as was noted
in \cite{lisaretiere}. This peculiar behavior is due to the constraint
that the collective flow is perpendicular to the constant $\tilde{r}$
surfaces. Thus, non-zero $v_2$ values even occur in the case where $\rho_a=0$.

The constraint that the collective flow is perpendicular to the surfaces of
constant $\tilde{r}$ is not justified from the perspective of
hydrodynamics. It is expected that the accelerations, which are driven by the
density gradients, are perpendicular to the surface of the ellipse, rather
than the velocities. For early times, before the shape has a chance to
react to the acceleration, the transverse acceleration and velocities should be
parallel to one another. But at later times, when $R_y/R_x$ approaches or
exceeds unity, the constraint enforces a rather awkward velocity profile. This
will be more apparent below where we consider the third parameterization.

\subsection{Hydrodynamically Inspired Blast Wave}

The second blast-wave model we will consider has the benefit of representing a
solution to non-relativistic hydrodynamics \cite{csorgo}. This parameterization
represents a solution to hydrodynamic equations in the limit that $P=nT$, and
particle number is conserved. Hydrodynamic can realize this form even if the
thermal motion is relativistic. The form also works for the simple
ultra-relativistic equation of state, $P\propto \epsilon$. However, this
parameterization does not represent an exact solution in the limit of high
collective velocities.

In this parameterization, the transverse profiles are Gaussian and the flow
velocities rise linearly from the origin.
\begin{eqnarray}
f&\propto&\frac{N}{(2\pi R_xR_y)}\exp\!\left(-n_q\frac{x^2}{R_x^2}
-n_q\frac{y^2}{R_y^2} \right)\exp(-E'/T), \nonumber\\
u_x&=&\alpha_x x,~u_y=\alpha_y y,
\end{eqnarray}
where $u_x$ and $u_y$ are the transverse relativistic velocities. As in the
previous model, it is assumed that the matter disintegrates at a fixed
temperature $T$. The factor $n_q$ in the exponential suggests that the
space-time structure of the Gaussian is driven by a position-dependent chemical
potential connected to the conserved quark number. In this parameterization the
elliptic flow does not depend on the radii and collective flow parameters
separately, but instead only on the combinations $\alpha_x R_x$ and
$\alpha_yR_y$. Since the collective velocities scale simply with the space-time
coordinates, it is straight-forward to express the spectra over integrals of
the collective velocities,
\begin{equation}
\frac{dN}{d^3p}\sim \int d\eta \cosh\eta \: u_\perp du_\perp d\phi_u
\exp\left[ - \frac{m_t}{T}\sqrt{1+u_\perp^2} + \frac{u_\perp}{T}
\cos(\phi_u-\phi_p)
- \frac{n_qu_x^2}{2\alpha_xR_x^2} - \frac{n_qu_y^2}{2\alpha_yR_y^2} \right] ,
\end{equation}
where $\phi_p$ refers to the direction of the momentum and $\phi_u$ is the
azimuthal direction of the transverse collective velocity $u_\perp$.

The integral can be performed analytically if the asymmetry is small.
Defining the parameters $\bar{u}$ and $\epsilon$,
\begin{eqnarray}
\bar{u}&=& \frac{1}{2}(\alpha_xR_x+\alpha_yR_y), \nonumber\\
\alpha_xR_x&=&\bar{u}(1+\epsilon),~\alpha_yR_y=\bar{u}(1-\epsilon),
\end{eqnarray}
leads to an expression for which the angular integrals can be performed
analytically after expanding to lowest power in $\epsilon$.
\begin{equation}
\frac{dN}{d^3p}\sim \int d\eta \cosh\eta \: u_\perp du_\perp
d\phi_u \left(1+n_q\epsilon \frac{u_\perp^2}{\bar{u}^2} \cos(2\phi_u)\right)
\exp\!\left[-\frac{m_t}{T}\sqrt{1+u_\perp^2} 
+ \frac{u_\perp}{T}\cos(\phi_u-\phi_p) - \frac{u_\perp^2}{2\bar{u}^2} \right].
\end{equation}
The calculation of $v_2$ involves an additional convolution with
$\cos 2\phi_p$, and gives
\begin{eqnarray}
v_2&=&n_q\epsilon \int u_\perp du_\perp (u_\perp^2/\bar{u}^2)
e^{-n_q u_\perp^2/2\bar{u}^2}K_1(z_E)I_2(z_p)/{\cal N}, \nonumber\\
{\cal N}&=&2\int u_\perp du_\perp
e^{-n_qu_\perp^2/2\bar{u}^2} K_1(z_E)I_0(z_p), \nonumber\\
z_E&=&(m_t/T)\sqrt{1+u_\perp^2},~~z_p= u_\perp p_t/T.
\end{eqnarray}

Results for this prescription are displayed in the third panel of
Fig. \ref{fig:blast} for the case of a 350 MeV constituent quark mass. The
temperature $T$ is considered the same as in the previous example and $\bar{u}$
is chosen to be 0.5. There is no peculiar negative elliptic flow from this
parameterization, and constituent scaling appears to be maintained to a
remarkably high level. Unlike the previous parameterization, we find no
sensitivity to the ratio $R_y/R_x$.

It can be shown that for small flow velocities and non-relativistic
temperatures, quark-number scaling becomes exact. However, this limit is far
from realized, and the apparent scaling is only approximate.

To test the sensitivity of the scaling to the Gaussian shape of the profile,
we repeated the calculation with an elliptic profile. For this calculation, the
collective transverse rapidities are again chosen to rise linearly with the
position, but the density is chosen to be uniform within an elliptic
region. Also, as in the case of the Gaussian source, $v_2$ is independent of
the spatial parameters. The results for the ellipse, shown in the lower panel
of Fig. \ref{fig:blast} are nearly identical to the Gaussian case. It thus
appears that the crucial factor for scaling is that the collective velocity
rises linearly with the position.

The fact that both the elliptic and Gaussian profiles are fairly successful at
reproducing quark-number scaling may be surprising given that the Gaussian
profile for the collective velocities has an explicit dependence on $n_q$,
while the elliptic profile does not. Since the Gaussian profile behaves as
$\exp(-n_qu^2/2\bar{u}^2)$, the average collective flow decreases as
$1/\sqrt{n_q}$.  Lower collective flows result in lower values for
$v_2$. However, since the distortions scale linearly with the density, $\sim
(1+n_q\epsilon\cos 2\phi_u)$, the two effects somewhat cancel one another to
some extent, and one obtains values of $v_2$ which depend mainly on
$p_t$. Since this dependence is roughly linear, scaling both the $x$ and $y$
axes by $n_q$, or any other constant, still results in a semi-universal curve.

Hydrodynamics provides for many more possibilities than the Gaussian density
profile with linear velocity profiles as described above. A strong first-order
phase transition will result in shock waves \cite{GulMat,Rischke}, which can
strongly violate the assumption of linear velocity profiles.  Additionally, the
temperatures might not be independent of the position. Nonetheless, most full
hydrodynamic solutions do tend to approach semi-linear forms by the time
breakup is approached \cite{huovinen}.

\subsection{Sensitivity to Mass}

In addition to the assumption that the effective volume referred to in
Eq. (\ref{eq:generalscaling}) was independent of $\phi$, the scaling of $v_2$
with quark number requires that for a particle $C$ composed of particles $a$
and $b$,
\begin{equation}
\label{eq:fproduct}
f_C(M_C{\bf u},x)=f_a(m_a{\bf u},x)f_b(m_b{\bf u},x),
\end{equation}
where ${\bf u} = \gamma {\bf v}$ is the relativistic velocity.  
For a thermal source where
\begin{equation}
\label{eq:thermalf}
f_i({\bf p},x)=\exp[-E'_i/T(x)+(\mu_i(x)/T(x)],
\end{equation}
Eq. (\ref{eq:fproduct}) is satisfied with a thermal distribution
if the three criteria are satisfied:
\begin{enumerate}
\item The temperatures describing $a$, $b$ and $C$ are the same.
\item The energies are conserved, $E_C(M_C)\approx
E_a(m_aP_C/(m_a+m_b))+E_b(m_bP_C/(m_a+m_b))$.
\item The chemical potential for $C$ is the sum of the constituent
chemical potentials, $\mu_C(x)/T(x)=\mu_a(x)/T(x)+\mu_b(x)/T(x)$.
\end{enumerate}
The first criteria requires only the assumption that the different species are
kinetically thermalized.

Since the energies are a mass multiplied by the gamma factor describing the
motion of the particles in the frame of the source at $x$, the second criteria
can be re-stated as a requirement that the masses sum, $M_C=m_a+m_b$. This is
equivalent to a requirement that there is no binding energy involved. However,
even if binding energy is involved, another simple scaling can be derived, as
discussed by Moln\'ar \cite{molnar_mscaling}. If the chemical potential is
ignored, Eq. (\ref{eq:thermalf}) may be written as
\begin{equation}
f_{\mu=0}(m_a{\bf u},x)=[f_{\mu=0}(m_0{\bf u},x)]^{m_i/m_0},
\end{equation}
which then implies that if one can ignore the contribution to $v_2$ from
the volume like term,
\begin{equation}
v_2^{(C)}=\frac{M_C}{m_a}v_2^{(a)}=\frac{M_C}{m_b}v_2^{(b)}.
\end{equation}
For the case where the quark masses are the same, this results in
\begin{equation}
\label{eq:mscaling}
v_2^{(C)}({\bf V})=\frac{M_C}{m_q}v_2^{(q)}({\bf V}).
\end{equation}
The scaling with mass differs from the quark-number scaling in two ways as can
be seen by considering the case of pions and protons. In quark number scaling
the proton will have 50\% more $v_2$ than a pion, while the difference will be
approximately a factor of seven for the case of mass-number scaling. Secondly,
for coalescence arguments based on conservation of momentum in the lab frame,
various particles are to be compared at the same value of ${\bf p}/n_q$,
whereas for mass-number scaling, the same velocity is to be compared, i.e., at
the same value of ${\bf p}/m$. Thus, for the case of mass-number scaling, a
universal curve will be obtained by plotting $v_2/m$ vs. $p_t/m$. Experimental
results seem to more closely follow quark-number scaling.

\begin{figure}
\centerline{\includegraphics[width=0.4\textwidth]{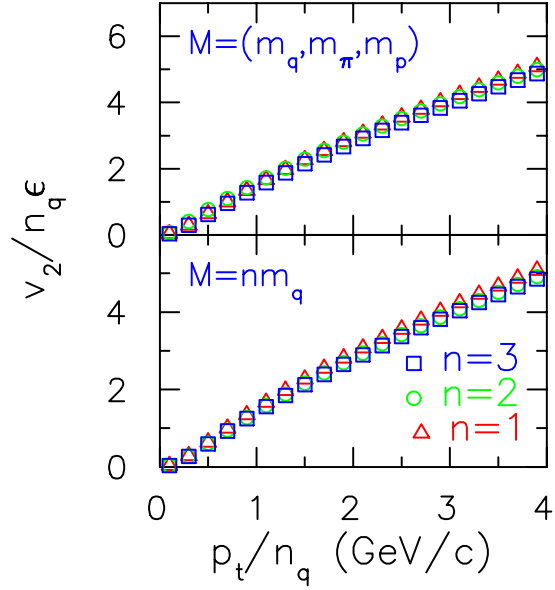}}
\caption{\label{fig:massdependence}
Scaled values for elliptic flow as a function of scaled 
transverse momentum for coalesced composite objects of 
$n_q=1,2,3$ quarks in the hydrodynamically inspired blast-wave model.
Lower panel: Composite objects have masses $M_C=n_qm_q$ with a 
constituent quark mass of $m_q=350$ MeV. 
Upper panel: Composite objects have masses $m_q$, pion mass, proton mass for 
$n_q=1,2,3$ respectively.
The values of $v_2$ are affected very little for $p_t/n_q \geq 1$ GeV/c. }
\end{figure}

In order to understand the degree to which mass effect destroy the scaling of
$v_2$ with quark-number, we repeated the calculations of the previous
subsection for the hydrodynamically inspired blast-wave model. Instead of using
hadronic masses equal to $M_c=350n_q$ MeV, we use the pion mass for the two
quark state and the proton mass for the three quark state. The comparison of
the two choices for masses is displayed in
Fig. \ref{fig:massdependence}. Remarkably little variation is observed for the
different curves. We also found no observable changes for the shell-profile
blast-wave calculations with similar mass modifications (not shown here). A
significant sensitivity to the mass is however observed if one focuses at low
$p_t$ below 500 MeV/c. The sensitivity is also more apparent for lower
collective velocities.

The third criteria for implementation of the thermal expression, i.e. the
addition of chemical potentials, implicitly requires that there are conserved
charges or numbers with $Q_C=Q_a+Q_b$. If quark number is conserved in the
hadronization process, the chemical potential can be associated with the net
number of quarks and anti-quarks. This is analogous to the implicit assumption
in coalescence that the quark content is unaltered by hadronization. Whereas
this assumption is certainly reasonable for the coalescence of protons into
deuterons, it can be questionable for the case of quark coalescence since quark
anti-quark pairs can be copiously created in hadronization as expected either
from entropy arguments or from dynamic pictures that involve string breaking.
However, for high $p_t$ mesons, it does seem reasonable to neglect
fragmentation processes, since such processes involve a color flux tube to form
between the outgoing quark and the thermal medium. This flux tube lowers the
momentum of the outgoing quark by roughly one unit, and in an exponential
spectrum, can be neglected at values of $p_t$ well above the temperature. The
breaking of such strings should be important for observables concerning
particles with $p_t<1.0$ GeV/c. One can make strong arguments that
hadronization for quarks below 1 GeV is of much different character than the
hadronization at higher values \cite{muellergroup}.

\section{Microscopic Simulations}
\label{sec:gromit}

More sophisticated models of heavy ion collisions, such as hydrodynamic or
Boltzmann descriptions, provide more complicated forms for the final phase
space density. By analyzing the results of these numerical models, one can gain
insight into the validity of the assumptions of blast-wave
parameterizations. Unfortunately, there is not yet a ``standard model'' for the
evolution of a RHIC collision. Hydrodynamic models using equations-of-state
based on lattice gauge theory tend to significantly over-predict the lifetime
of the collision \cite{starcor,phenixcor,Heinz}.  Boltzmann approaches which
effectively represent stiff equations of state tend to somewhat under-predict
source sizes \cite{molnarhbt}. Since the issue of the space-time development of
the collision remains uncertain at the 50\% level, we will instead focus on a
simplified Boltzmann description. By adjusting the cross sections and the
initial conditions, we intend to better understand how and why $v_2(y_t)$ can
depend on $n_q$.

For the first calculation, we consider a system of 500 constituent quarks per
unit rapidity of mass 350 MeV and initially thermalized at a temperature of 275
MeV. The particles are then allowed to collide with simple s-wave cross
sections using the code GROMIT \cite{gromit} to solve the Boltzmann
equation. The initial density distribution in the transverse directions are
set by either a Gaussian distribution, or by an ellipse with constant
density. The initial sizes for the Gaussian profiles are chosen to be $R_x=2$
fm and $R_y=2.6$ fm, while for the ellipse, the radii are 4 and 5.2 fm. The
longitudinal distributions are initialized according to boost invariant
criteria, $v_z=z/t$. Boost-invariant constraints are maintained by cyclic
boundary conditions along the $z$ axis.

\begin{figure}
\centerline{\includegraphics[width=0.45\textwidth]{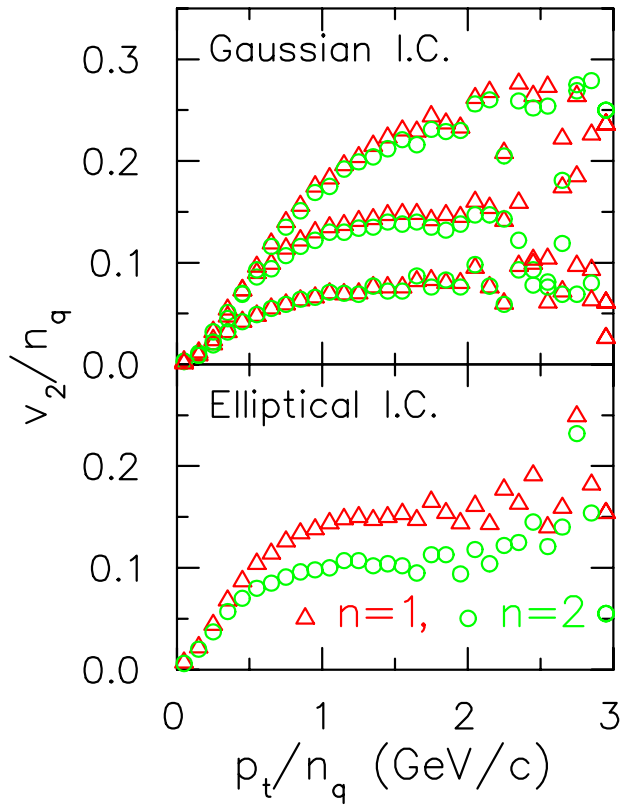}}
\caption{\label{fig:gromit} Scaled values for elliptic flow $v_2/n_q$ as a
function of scaled transverse momentum $p_t/n_q$ from Boltzmann transport
calculation.  Upper panel: The initial state evolved from Gaussian density
profile with elastic scattering cross sections of $\sigma= 5, 20$ and 80 mb
(bottom to up).  Nearly perfect scaling with quark number is exhibited.  Lower
panel: The initial state evolved from elliptic density profile with a sharp
cutoff. $v_2$ significantly violate quark-number scaling. }
\end{figure}

The upper panel of Fig. \ref{fig:gromit} shows calculated values of $v_2$ from
simulations using three cross sections, 5 mb, 20 mb and 80 mb. In addition to
the $v_2$ for constituent quarks ($n_q=1$), $v_2$ for coalesced mesons
($n_q=2$) are also shown. Meson coalescence is calculated by sorting the quarks
into bins according to their transverse momentum $p_t$ and azimuthal angle
$\phi_p$, then coalescing them with a weight proportional to $\exp(-\delta
r^2/2r_c^2)$, where $\delta r$ is the separation of the two quarks as measured
in the two-particle rest frame. The coalescence distance $r_c$ is chosen to be
1 fm. Constituent number scaling appears to be satisfied to high precision for
all three cross sections. This is probably due to the fact that the initial
conditions and equation of state satisfy the conditions for the scaling
hydrodynamic solution discussed in the previous section, which in the limit of
small flow velocities can be shown to lead to quark-number scaling.

Calculations based on an initial elliptic profile with sharp cutoff are
displayed in the lower panel of Fig. \ref{fig:gromit}. Aside from the density
profile, the calculation was the same as was shown for the Gaussian profile
with a cross section of 20 mb. In this case, quark-number scaling is
significantly violated. This failure can be attributed to the shock wave which
develops from the discontinuity in the density. From the perspective of
hydrodynamics, there is no pressure gradient except at the surface. This forces
the development of a shock wave. Emission is then more surface-like and
less volume-like, which is more suggestive of the illustration on the right
hand side of Fig. \ref{fig:effvolcartoon}. For surface-like emission, one
expects the asymmetry to owe itself to the disparity in the effective surface
areas for emitting in- and out-of-plane, rather than to asymmetries in the
respective phase space densities. For equations of state that incorporate a
large latent heat, shock waves also develop since there is a large region of
mixed phase for which there is no pressure gradient and therefore no transverse
acceleration \cite{GulMat,Rischke}. One may also expect to observe violations
from scaling in these instances. It will be interesting to analyze hydrodynamic
calculations to see whether the presence of a large latent heat will, indeed,
manifest itself in the behavior of $v_2$.

\section{Summary}
\label{sec:summary}

Quark-number scaling for elliptic flow has been advocated as a signal for
quark recombination. The two principal goals of this study are to identify
precise conditions for quark number scaling, and to determine the degree to
which thermal, coalescence, and microscopic models may realize these
conditions.

Two conditions are necessary for realizing quark-number scaling. First, the
phase space density for a composite object must equal the product of the phase
space densities of the constituent quarks. Second, the effective volume for
quarks that have a given momentum must be independent of azimuthal
direction. It will be much more difficult to determine the validity of the 
latter condition since the effective volume is determined both by details of 
the spatial source size and by the collective flow and temperature.

The requirement of factorization is manifestly satisfied by coalescence
models. It is approximately satisfied by thermal models, and becomes exactly
satisfied in the limit that the masses of the coalesced hadron equal the
product of $n_q$ and the constituent quark mass. However, we never observe a
significant dependence to the behavior of $v_2$ as we varied the masses of the
final hadrons for $p_t/n_q$ > 1 GeV/c, the range of interest for quark
recombination arguments. It should be emphasized, that in a thermal model, the
only means to have non-uniform densities are by varying the temperature and
chemical potential. If non-uniform chemical potentials are responsible for the
density profile, thermal models implicitly assume that charge must be conserved
by hadronization. For the hydrodynamically inspired blast-wave with a Gaussian
profile, the temperature is kept constant, and the density is driven by
non-uniformities in the chemical potential. Quark-number scaling for this model
is then dependent on the baryons having 3/2 the charge of a meson, although
this sensitivity is weak due to the linearity of $v_2$ vs. $p_t$.

The models invoked in this study are schematic by nature, but they provide
several important informations regarding the interplay of coalescence and
elliptic flow. It does appear that linear velocity profiles with sudden
time-like break-up surfaces promote constituent-number scaling criteria of
Eq. (\ref{eq:v2scaling}). This is verified by both blast-wave and microscopic
modeling. Since linear velocity profiles tend to develop, except in the
presence of strong shocks, Eq. (\ref{eq:v2scaling}) should be valid unless the
initial density profile is abnormally sharp, or if the equation of state has a
strong first-order phase transition.

Furthermore, the scaling condition can be tested by measuring the source-size
as a function of the azimuthal volume. This can be done either with
correlations or with coalescence ratios. If the source-size volumes at a given
$p_t$ are independent of the azimuthal angle, it follows that
constituent-number scaling of $v_2$ should be satisfied if the factorization
criteria are met. Perhaps the easiest way to test this will be to estimate the
$d/p^2$ ratio for deuterons and protons at the same transverse rapidity. This
ratio is inversely proportional to the source size volume, so if
constituent-number scaling is valid, it should be independent of the azimuthal
angle. In other words, the $v_2$ for deuterons will then be twice $v_2$ for
protons.

We have explored numerous models for the satisfaction of quark-number scaling:
\begin{enumerate}
\item Four different distortions of a blast-wave with an elliptic-shell
geometry.
\item The blast-wave model of Retiere and Lisa, with both in-plane and
out-of-plane extended profiles.
\item A blast-wave with simple linear behavior for the collective velocity and
either a Gaussian or sharp elliptic profiles for the density.
\item A solution of the Boltzmann transport equation with a coalescence
prescription for combining quarks into hadrons. Both Gaussian and sharp
elliptic profiles are used for the initial densities.
\end{enumerate}
In each of these prescriptions, the factorization condition is explicitly
satisfied. Thus, the satisfaction of quark-number scaling depend solely on
whether the effective volumes are $\phi$-independent. Unfortunately, it is not
algebraically possible to show that any of the models explicitly satisfy
quark-number scaling, except for the blast-wave with linear velocity gradients
and a Gaussian profile in the limit of small expansion velocities and
non-relativistic temperatures. However, an empirical pattern does seems to
emerge: Linear velocity gradients and sudden volume-like disintegrations seem
to closely satisfy quark-number scaling. Hydrodynamic evolutions tend to
approach this geometry, except in the case of surface-like emission, which will
be favored if shock waves are to develop. Shock waves are expected if the
initial density has a sharp fall-off, or if there is a large latent heat which
will lower the interior pressure of the fireball to give a shock wave the
opportunity to develop.

Quark recombination is synonymous with satisfaction of the first condition,
factorization of the phase space density of the combined object. This is
because factorization assumes that two-particle densities are products of
single-particle densities, i.e., jet-like correlations are lost. However, since
factorization represents only one of the two conditions for quark-number
scaling of $v_2$, there is always some ambiguity in concluding that both
conditions are satisfied when one observes the scaling. For instance, it could
be that neither condition is satisfied, but that the failures to meet both
conditions conspired in some way to mimic the satisfaction of both criteria. Of
course, a stronger case can be made if scaling is observed for a large number
of cases. However, one must also remember that if the various cases involve
hadrons with different cross sections, i.e., strange vs. non-strange hadrons,
then one may expect violations of $v_2$. This can be understood by viewing
Fig. \ref{fig:gromit} where the overall value of $v_2$ is shown to be strongly
sensitive to the cross section.

Our investigations of both model results and of the underlying theory shows
that observing quark-number scaling of $v_2$ indeed suggests that quark
recombination may be at hand, but there are quite a large number of caveats
that prevent one from declaring this as a stand-alone signature. However, it is
clear that $v_2$ is, at the very least, very sensitive to the evolution,
break-up mechanism and hadronization of the fire-ball. Thus, the details of the
behavior of $v_2$ can be used to disqualify a large number of potential models,
even if pointing to the root cause of the disqualifying attributes might be
somewhat ambiguous.

We conclude by recommending that source-size measurements at intermediate $p_t$
can significantly clarify some of the issues surrounding quark recombination.
First, interferometric or coalescence measurements may determine the source
size as a function of $p_t$ and $\phi$. If source sizes are seen to be
independent of the azimuthal angle, it should support the statement that
quark-number scaling demonstrates that phase space densities factorize. More
importantly, a simpler and more direct point can be made about such source-size
measurements. If all particles in the intermediate-$p_t$ region with the same
azimuthal angle originate from the same jet, the source size from correlation
or coalescence analyses will be identical to that expected from $pp$ collisions
of $\sim 1$ fm. If the emissions are independent, and factorization is valid,
the source sizes will reflect the entire emitting region of such
particles. Even if emission of high energy particles is confined to the
surface during the first few fm/c, the effective volumes will be on the order
of tens of cubic fm. The real value probably lies somewhere in between. Since
jet-like correlations are observed at these $p_t$ region
\cite{expjetcorrelations}, the assumption of completely independent sources is
probably not viable and more detailed models which incorporate jet-like
correlations into coalescence pictures are currently being developed
\cite{friesjetcorrelations,molnar2004}. A measurement of the source volume will
then provide crucial quantitative information to decide the relative role of
recombination.

\begin{acknowledgments}
This work was supported by the U.S. National Science Foundation, Grant No.
PHY-02-45009 and by the U.S. Department of Energy, Grant No. DE-FG02-03ER41259.
\end{acknowledgments}

\end{document}